\documentclass[twoside,12pt]{article}

\usepackage{epsfig}

\begin{document}

\pagestyle{myheadings}
\markboth{AADEBUG 2000}{Model-based Debugging of Java Programs}

\newcommand{\Forall}{\mbox{$\forall\,$}}
\newcommand{\Exists}{\mbox{$\exists\,$}}
\newcommand{\And}{\mbox{$\,\wedge\,$}}
\newcommand{\Or}{\mbox{$\,\vee\,$}}
\newcommand{\impl}{\mbox{$\,\supset\,$}}
\newcommand{\union}{\mbox{$\,\cup\,$}}
\def\Nat{\rm I\hspace*{-0.2em}N}
\newenvironment{literatur}%
{\begin{quote}\leftmargin4cm\begin{flushright} \sf }%
{\end{flushright}\end{quote}}

\newcommand{\jade}{{\sf Jade~}}
\newcommand{\p}{\!::\!}
\newcommand{\F}{{\cal F}}


\newenvironment{programlist}%
{\small\begin{tabbing} \quad \=\quad \=\quad \=\quad \=\quad \=\quad \= \kill}%
{\end{tabbing}}
\newcommand{\keyw}[1]{{\textbf {#1}}}

\title{Extended Abstract -- Model-Based Debugging of Java Programs\thanks{This
work was partially supported by the Austrian Science Fund project
P12344-INF and project N~Z29-INF.}
\footnote{In M. Ducass\'e (ed), proceedings of the Fourth
International Workshop on Automated Debugging (AADEBUG 2000), August
2000, Munich. COmputer Research Repository (http://www.acm.org/corr/),
cs.SE/0011027; whole proceedings: cs.SE/0010035.}}

\author{Cristinel Mateis \and Markus Stumptner \and Dominik Wieland
\and Franz Wotawa \thanks{Authors are listed in alphabetical
order.} \\
Technische Universit{\"a}t Wien \\
Institut f{\"u}r Informationssysteme \\
Database and Artificial Intelligence Group \\
Favoritenstra{\ss}e 9--11, A-1040 Vienna, Austria \\
\{mateis,mst,wieland,wotawa\}@dbai.tuwien.ac.at \\
http://www.dbai.tuwien.ac.at}

\date{}

\maketitle

\section{Introduction}

Model-based reasoning is a central concept in current research into
intelligent diagnostic systems.  It is based on the assumption that
sources of incorrect behavior in technical devices can be located and
identified via the existence of a model describing the basic
properties of components of a certain application domain.  When
actual data concerning the misbehavior of a system composed from such
components is available, a domain-independent diagnosis engine can be
used to infer which parts of the system contribute to the observed
behavior.  Model-based Diagnosis provides a set of proven algorithms
and methods for searching faults~\cite{rei87} and identifying points
of measurement~\cite{kle87short}.

This paper describes the application of the model-based approach to
the debugging of Java programs written in a subset of Java.  We show
how a simple dependency model can be derived from a program,
demonstrate the use of the model for debugging and reducing the
required user interactions, give a comparison of the functional
dependency model with program slicing~\cite{wei84}, and finally
discuss some current research issues.

\section{Model-Based Diagnosis}

The model-based approach is based on the notion of providing a
representation of the correct behavior of a technical system.
By describing the structure of a system and the
function of its components, it is possible to ask for the
reasons why the desired behavior was not achieved.  In the diagnosis
community, the model-based approach has achieved wide recognition due
to its advantages:

\begin{itemize}
\item once an adequate model has been
developed for a particular domain, it can be used to diagnose
different actual systems of that domain
\item the model can
be used to search for single or multiple faults in the system without
alteration
\item different diagnosis algorithms can be used for a given model
\item the existence of a clear formal basis for judging and computing diagnoses
\end{itemize} 

Using the standard consistency-based view as defined by
Reiter~\cite{rei87}, a diagnosis system can be seen formally as a
tuple $(SD,COMP)$ where $SD$ is a logical theory sentence modeling the
behavior of the given system (in our case the program to be debugged),
and $COMP$ a set of components, i.e., statements. A diagnosis system
together with a set of observations $OBS$, i.e., a test-case, forms a
diagnosis problem. A diagnosis $\Delta$, i.e., a bug candidate, is a
subset of $COMP$, with the property that the assumption that all
statements in $\Delta$ are incorrect, and the rest of the statements
is correct, should be consistent with $SD$ and $OBS$. Formally,
$\Delta$ is a diagnosis iff $SD \cup OBS \cup \{ \neg AB(C) | C \in
COMP \setminus \Delta\} \cup \{ AB(C) | C \in \Delta \}$ is
consistent. A component not working as expected, i.e., a statement
containing a bug, is represented by the predicate $AB(C)$. 

The basis for this is that an incorrect output value (where the
incorrectness can be observed directly or derived from observations
of other signals) cannot be produced by a correctly functioning
component with correct inputs.  Therefore, to make a system with
observed incorrect behavior consistent with the description and avoid
a contradiction, some subset of its components must be assumed to work
incorrectly.  In practical terms, one is interested in finding minimal
diagnoses, i.e., a minimal set of components whose malfunction
explains the misbehavior of the system (otherwise, one could explain
every error by simply assuming every component to be malfunctioning).
Basic properties of the approach as well as algorithms for efficient
computation of diagnoses are described in~\cite{rei87}.

Starting from straightforward work that used a logic program directly
as system descriptions~\cite{con93,bond94a}, in the last years the use
of model-based reasoning (MBR) for debugging of software has been
examined in a wider context~\cite{trav94,fsw99,sw99a,msw99,msw00}.
All of the approaches have in common that they use a model derived
from a program for locating (or, rarely, correcting) a bug. They
differ in the considered programming language (ranging from purely
logical languages to hardware description languages -- in particular
VHDL~\cite{fsw99}, functional, and finally imperative languages), and
the type of model (qualitative~\cite{trav94}, dependency-, or
value-based models). The purpose of previous research was to
show the applicability of MBR in the software domain by introducing
models, often for special purpose languages. Our current work deals
with the extension and application of these principles to a mainstream
language (Java). This paper presents first results of an
implemented debugger prototype using different example programs. The
{\sf JADE} debugger currently implements a functional dependency model
that extends our earlier work~\cite{msw99,msw00}. The granularity of
the debugger, i.e., the elements of a program that are considered to
be faulty or not, is currently set to the statement level (instead of
individual expression level) for efficiency reasons.

\begin{figure}[t]
\begin{center}
\epsfig{file=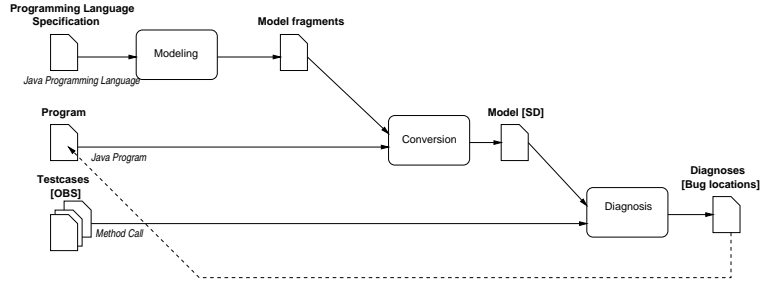,
width=10cm}
\end{center}
\caption{Model-based Diagnosis for Debugging}
\label{fig:mbddeb}
\end{figure}

The {\sf JADE} debugger combines the standard operation modes of
diagnosis systems and standard debuggers. First, the program is
converted into the dependency representation which is compiled into a
logical model. Once the program has been executed actual
observations of its behavior can be provided. This behavior
together with the logical model is used by the diagnosis engine
to compute bug candidates that map back to positions (and statements)
within the program to be debugged (see Figure~\ref{fig:mbddeb}).

\begin{figure}[t]
\begin{center}
\begin{minipage}{5cm}
\begin{programlist}
1. \> \keyw{class} SWExamples \{ \\
2. \> \> \keyw{public static void} test(int a,b,c,d,e) \{ \\
3. \> \> \> int f,g,s1,s2,s3; \\
4. \> \> \> s1=a*c; \\
5. \> \> \> s2=b*d; \\
6. \> \> \> s3=c*e; \\
7. \> \> \> f=s1+s2; \\
8. \> \> \> g=s2+s3; \\
9. \> \> \} \\
10. \> \}
\end{programlist}
\begin{center}
(a) Source code
\end{center}
\end{minipage}~~~~
\begin{minipage}{5cm}
\begin{center}
{\small
\begin{tabular}{ll}
\hline
Line & Environment \\
\hline
   & test(3,2,2,3,3) \\
2. & $a_{test}=3$, $b_{test}=2$, $c_{test}=2$,\\
   & $d_{test}=3$,$e_{test}=3$ \\
3. & \\
4. & $s1_{test}=6$ \\
5. & $s2_{test}=6$ \\
6. & $s3_{test}=6$ \\
7. & $f_{test}=12$ \\
8. & $g_{test}=12$ \\
9. & \\
10.& $test(3,2,2,3,3)=void$ \\
\hline
\end{tabular}
}
\end{center}
\begin{center}
(b) Evaluation Trace for test(3,2,2,3,3)
\end{center}
\end{minipage}
\end{center}
\caption{A simple Java method}
\label{fig:d74swexample}
\end{figure}

\section{Modeling for Debugging}

We first give an overview of computing the dependencies for Java
programs and then describe the derivation of the system description
which is used for diagnosis. 

\paragraph{Dependencies}
For the sake of brevity we omit the discussion of object-oriented
features such as dynamic binding in this paper and concentrate on the
"basic" imperative features of the language.  A more detailed
technical description of the basic idea behind the conversion
algorithm (excluding external side effects and method calls) can be
found in~\cite{msw99}, and a discussion about references and side
effects in~\cite{msw00}. In order to compute dependencies, we must
consider that the variables occurring in the methods change their
values during program execution. This is handled by assigning a unique
index to all locations where a variable occurs as target of an
assignment (we refer to each such location as an {\em occurrence} of
the variable.  It is the various variable occurrences that
dependencies are computed for.

Let $x$ and $y$ be indexed variable occurrences of a given method $m$.
We say that $x$ {\em depends on} $y$ iff the value of $y$ determines
the value of $x$ for at least one input vector.  This definition is
based on earlier work on dependencies in software debugging, e.g.,
~\cite{korel88,MoW90}. Beside debugging dependencies are used for
verification (see~\cite{jackson95}).  Formally, we define a functional
dependency as a pair $(x,M_x)$, where $x$ is a variable occurrence and
$M_x$ is a set of variable occurrences such that $x$ depends on every
$y\in M_x$. We can now compute all functional dependencies of a
particular statement by determining the functional dependencies of all
variables used within the statement.

Whereas the variables correspond to the ports of traditional diagnosis
components, the natural choice for components given the abstract
nature of the dependency-based representation are statements.  The set
of diagnosis components can be viewed as a diagnosis system where the
connections are formed by the variable occurrences inside the
components, i.e., components $c_i$ and $c_j$ are connected iff one
component establishes the functional dependency $(v_i,M)$ and the
other, $(w_j,\{\ldots,v_i,\ldots\})$.  Variable occurrences $v_0$ are
inputs of the whole diagnosis system.  A variable occurrence $v_i$ is
an output iff there is no other occurrence $v_j$ such that $j > i$.
Since during conversion indices are always increased, the resulting
diagnosis system, i.e., the graph representation, is acyclic.

Computing functional dependencies for Java programs requires compiling
each method declared for a class. A method $m$ of a Java program is
converted by sequentially converting its statements into diagnosis
components.  

We illustrate the computation of functional dependencies using the
example of Figure~\ref{fig:d74swexample}. The functional dependencies
for {\tt s1=a*c;} are $\{(s1_2,\{a_1,c_1\})\}$ because the value of
{\tt s1} is given by the product of the values of {\tt a} and {\tt
c}. In summary we obtain the following dependency sets (indices are
ignored) for the 5 statements: $fd(C_4) = \{ (s1,\{a,c\}) \}$,
$fd(C_5) = \{ (s2,\{b,d\}) \}$, $fd(C_6) = \{ (s3,\{c,e\}) \}$,
$fd(C_7) = \{ (f,\{s1,s2\}) \}$, $fd(C_8) = \{ (g,\{s2,s3\}) \}$ where
$C_i$ denotes the statement in line $i$.

\paragraph{The System Description}
After computing all dependencies, we map them to a logical
representation, which can be directly used for model-based debugging.
For this purpose we assume that the statements are given as a set
$COMP$, and that for all statements, the functional dependencies are
defined. The set of functional dependencies for a statement $st$ is
written as $fd(st)$.

Functional dependencies describe behavior implicitly by describing
influences between variables. Instead of speaking about real values,
we can only speak about whether a value $v$ is correct (written as
$ok(v)$) or not (written $nok(v)$).  We can further write that if a
statement $s$ is assumed to be correct (i.e., $\neg AB(s)$ holds) and
all input variables have a correct value then the value of variables
used as target in an assignment statement must be correct. Formally,
the system description is given by:
$$\Forall_{(o,M) \in fd(C)} \left[ \neg AB(C) \And \bigwedge_{x \in M}
ok(x) \rightarrow ok(o) \right] \in SD$$
where $C\in COMP$ is a statement. In addition, we know that it is
impossible that a variable value is known to be correct and incorrect
at the same time. Therefore, we have to add the rule $ok(v) \And
nok(v) \rightarrow \bot$ to the model $SD$, for each variable
occurrence $v$ in the program. 
The described model can be used together with a standard MBD algorithm
for computing bug locations.

For software debugging, the observations required for diagnosis are
given by the specified behavior, in our case the expected input/output
vectors. By comparing the specified output with the computed output,
we can classify the correctness of variables. Variables $v$ that are
assumed to have the correct value lead to the observation $ok(v)$.
Variables with an incorrect value are represented by $nok(v)$.

In Figure~\ref{fig:d74swexample}(b) the evaluation
trace for the call {\em test(3,2,2,3,3)} for the Java program given in 
Figure~\ref{fig:d74swexample}(a) is given. The trace only
presents the lines of code which are involved in the current
evaluation, and the new environments created. To distinguish different
local variables they are indexed with the name of the method where
they are declared. In this case there is no return value. From the
dependencies computed above, we get the logical model $SD$:

\begin{center}
\small
\begin{tabular}{cc}
$\neg AB(C_4) \And ok(a) \And ok(c) \rightarrow ok(s1)$ &
$\neg AB(C_5) \And ok(b) \And ok(d) \rightarrow ok(s2)$ \\
$\neg AB(C_6) \And ok(c) \And ok(e) \rightarrow ok(s3)$ &
$\neg AB(C_7) \And ok(s1) \And ok(s2) \rightarrow ok(f)$ \\
$\neg AB(C_8) \And ok(s2) \And ok(s3) \rightarrow ok(g)$ &
\end{tabular}

\begin{tabular}{cccc}
$ok(a) \And nok(a) \rightarrow \bot$ &
$ok(b) \And nok(b) \rightarrow \bot$ &
... &
$ok(s3) \And nok(s3) \rightarrow \bot$\\
\end{tabular}
\end{center}

In this example we assume that the method call {\em test(3,2,2,3,3)}
should lead to values {\em f=12} and {\em g=0}, i.e., that line 8
should be {\em g=s2-s3} instead of {\em g=s2+s3}. For this case, we get
observations $OBS$:

$$ ok(a) \And ok(b) \And ok(c) \And ok(d) \And ok(e) \And ok(f) \And nok(g)$$

Using $SD \cup OBS$ we get 3 diagnoses, each pinpointing a single
possible bug location: $\{C_5\}, \{C_6\}, \{C_8\}$. The other
statements can be ignored in this case.  Using the
measurement selection algorithm from~\cite{kle87short} we can compute the
optimal next question to be presented to the user in order to
distinguish between the 3 candidates.

\section{Diagnosing with the Dependency Model}

The {\sf JADE} debugger with dependency model can be proven to be complete
with regard to bugs that do not alter the dependency structure of the
program, since all statements that may cause a wrong value are
considered and therefore are diagnosis candidates. However,
discrimination capability can be low.  Consider the example program
from Figure~\ref{fig:d74swexample}, together with the specified values
$f=12$ and $g=0$. In this case, the debugger returns the candidate
$\{C5\}$ which could be eliminated when using a value-based model. Now
assume $C5$ is incorrect, all other statements are correct, and apply
the test case from Figure~\ref{fig:d74swexample}. From $C4$ and $C6$
and the input values we derive $s1_{test}=s3_{test}=6$. Using these
values together with $C7$ and $f_{test}=12$, we get $s2_{test}=6$. Now
using this value together with the assumption $C8$ is correct leads to
$g_{test}=12$, contradicting our specified value $g_{test}=0$. Hence,
$\{C5\}$ is not a diagnosis w.r.t the value-based model, illustrating
the (unsurprising) fact that a model based purely on dependencies is
too weak to discriminate between all possible program errors.

Concerning performance, for smaller programs and interactive
debugger use, diagnosis times should be in the single second range
although longer times are acceptable for very large programs.  It is
obvious that searching for all single bugs using our model is
restricted to $O(n^2)$, where $n$ denotes the number of diagnosis
components, i.e., in our case, statements. Using empirical results
from~\cite{nej97} we can expect that computing all single bugs for
Java methods with several hundred statements should be done in less
than 1 second, a result that is consistent with the experience from the
VHDL domain~\cite{fsw99}.

Like program slicing~\cite{wei84}, our dependency model is based
on static analysis of the code, i.e., it is computed using the program
structure and does not use the runtime program behavior for fault
localization. A program slice is defined as the part of a program
possibly influencing the value of given variables and not occurring
within the program after a given position. The slice for our running
example for variables $\{g\}$ and position 8 comprises the lines 8, 6,
and 5. This result is equal to the one obtained by our dependency
model and the question arises about the differences between both
approaches. Our dependency model is more hierarchically organized,
e.g., formally a conditional statement is viewed as a single diagnosis
component and sub-divided only after being identified as faulty. To
allow a comparison of slicing with MBD using the functional dependency
model, we assume an appropriate mapping. In this respect we obtain
similar results from both techniques except in the cases where several
variables have a faulty value after program execution. In this
situation the model-based approach tries to minimize the source of the
misbehavior leading to fewer solutions, while slicing does not. Given
the example, assume that line 5 is faulty, leading to wrong values for
variables $f$ and $g$. The slice for $f$ and $g$ is the whole program,
while our dependency model would deliver only line 5 as single fault
candidate.

\begin{table*}[t]
\begin{center}
\scriptsize
\begin{tabular}{|l|l|l|l|l|l|l|l|l|l|l|} \hline \hline
Test&	Method&			Lines&	Error&	\multicolumn{6}{c}{Interactions}& \\ \hline
&	&			&	&	Setup&	Query&	Loop&	Exprs.&	Iter.&	Total&	Total 2 \\ \hline \hline
1&	$adder\_f1$&		17&	4&	1&	2&	0&	1&	0&	4&	3 \\
2&	$adder\_f2$&		17&	4&	1&	2&	0&	1&	0&	4&	3 \\
3&	$adder\_f3$&		17&	7&	1&	2&	0&	1&	0&	4&	3 \\ 
4&	$adder\_f4$&		17&	7&	1&	2&	0&	1&	0&	4&	3 \\ 
5&	$adder\_f5$&		17&	7&	1&	2&	0&	1&	0&	4&	3 \\ 
6&	$adder\_f6$&		17&	12&	1&	4&	0&	1&	0&	6&	5 \\
7&	$adder\_f7$&		17&	12&	1&	4&	0&	1&	0&	6&	5 \\
8&	$adder\_f8$&		17&	11&	1&	4&	0&	1&	0&	6&	5 \\
9&	$adder\_f9$&		17&	11&	1&	6&	0&	1&	0&	8&	7 \\
10&	$adder\_f10$&		17&	14&	1&	4&	0&	1&	0&	6&	5 \\
11&	$adder\_f11$&		17&	14&	1&	4&	0&	1&	0&	6&	5 \\
12&	$adder\_f12$&		17&	16&	1&	4&	0&	1&	0&	6&	5 \\
13&	$adder\_f13$&		17&	12&	1&	4&	0&	1&	0&	6&	5 \\
14&	$adder\_f14$&		17&	9&	1&	4&	0&	1&	0&	6&	5 \\ \hline
15&	$library\_f1$&		30&	28&	1&	5&	2&	1&	1&	10&	6 \\ \hline
16&	$bubbleSort\_f1$&	10&	4&	1&	2&	2&	1&	1&	7&	3 \\
17&	$bubbleSort\_f2$&	10&	7&	1&	5&	3&	1&	2&	12&	6 \\
18&	$insertionSort\_f$&	19&	4&	1&	2&	2&	1&	1&	7&	3 \\
19&	$shellSort\_f1$&	14&	3&	1&	2&	1&	1&	1&	6&	3 \\
20&	$shellSort\_f2$&	14&	6&	1&	3&	3&	1&	2&	10&	4 \\
21&	$selectionSort\_f1$&	14&	1&	1&	1&	0&	1&	0&	3&	2 \\	
22&	$heapSort\_f1$&		11&	7&	1&	6&	1&	1&	1&	10&	7 \\
23&	$heapSort\_f2$&		11&	8&	1&	4&	1&	1&	1&	8&	5 \\
24&	$heapSort\_f3$&		11&	9&	1&	3&	1&	1&	1&	7&	4 \\ \hline 
Sum&	&			382&	217&	24&	81&	16&	24&	11&	156&	105 \\ 
Av.&	&			15.92&	\emph{9.042}&	1&	3.37&	0.67&	1&	0.46&	6.5&	\emph{4.37} \\ \hline \hline
\end{tabular}
\end{center}
\caption{Debugging results from Java examples}
\label{fig:results}
\end{table*}

\section{Empirical results}

The following experiments show the results of the {\sf JADE} debugger
using various Java methods from our example library, modified at
randomly selected statements: \\ {\bf Example 1:} The $adder$ method
implements a binary full adder mapping three inputs to two outputs. \\
{\bf Example 2:} The $library$ method is part of a small application
that creates a sample library and then computes the author who has
published the most books of all authors whose books can be found in
the library (methods involving object-oriented language structures,
such as multiple objects, instance method calls, class \& instance
variables, etc...) \\ {\bf Example 3:} Sorting methods: various
sorting methods, providing the full complement of control statements:
loops, selection statements, and method calls.

Table~\ref{fig:results} shows (from left to right) the tested method
(in which a single error has been installed), the total number of
statements in each method, the index of the buggy statement within the
method (which can directly be used to compare the outcomes of the
\jade\ debugger tests with "manual" use of a debugger where the user
steps through the code sequentially until the erroneous line is
found), and finally the number of user interactions which are needed
to exactly locate the bug, classified by type: \\
{\bf Setup:} verify the system output connections, i.e. all output
variables of the method. \\ 
{\bf Query:} verify the value of a variable at a particular statement \\
{\bf Loop:} debug the condition of a loop or selection statement \\
{\bf Exprs:} find the smallest sub-expression of a particular
statement. This allows further debugging of method and constructor
calls. \\
{\bf Iter.:} determine the first loop iteration in which the error occurs 

The "Total" column from the right shows the total number of user
interactions, and the "Total2" column the sum of all setup and query
interactions.  The latter figure determines the debugger's performance
at statement level and can therefore be compared with Column 4.  These
two columns give a simple comparison of the user interactions needed
to find the exact location of a faulty statement with the two
different debugging strategies, i.e. model-based vs. unsupported
debugging.

Locating a bug in the $adder$ method with a traditional debugging tool
requires 10 user interactions versus 4.43 with the {\sf JADE} debugger.
This fairly drastic difference in favor of the model-based technique is
due to the simple block structure of the method with the model-based
debugger exploiting its knowledge about the underlying functional
dependency structure of the block.  The longer the block, the greater
the advantage.  For the same reason, the faulty version of the
$library$ method requires 28 user interactions in the traditional
approach compared to 6 using the {\sf JADE} debugger.
To locate a bug in a sorting algorithm a traditional debugger on
average needs 5.45 steps versus 4.12 for the {\sf JADE} debugger, the
advantage being less clear due to the complex control structures.

Overall, 24 tests are compiled in the table with 9.04 user interactions
on average required in the traditional case and 4.37 for the {\sf JADE}
debugger, clearly highlighting the potential abilities of model-based
debugging tools.

Note that as with techniques like program slicing, the bug location
efficiency of the model-based approach does not depend on the actual
location of the error in the code. The number of user interactions it
takes to find the bug is therefore much better assessable.

\section{Discussion and Conclusion}

One of the main advantages of the model-based approach is the ability
to incorporate multiple models in the same formal and computational
framework.  The dependency-based representation described in this
paper has the advantage of being simple and computationally efficient
to solve, thus providing an answer to the important question of
scalability of the approach (our use of a dependency based mechanism
in the VHDL domain was usable for very large programs~\cite{fsw99}).
Its disadvantage of low discrimination can be counteracted by
combining it with a more detailed (i.e., value-based) model, e.g., by
focusing in on smaller program parts isolated by the dependency model,
and diagnosing them with the more detailed model, which is currently
undergoing implementation~\cite{msw00}.  The value-based model, unlike
the dependency model, is dynamic and based on the evaluation trace,
similar to the manner in which Dynamic Program Slicing~\cite{korel97}
extends classical Program Slicing.  The value-based model records, in
a detailed manner, which program parts actually contribute to the
faulty behavior.  By propagating value assumptions forward and
backward via the standard diagnosis algorithms it provides better
discrimination. In~\cite{fritz91} an approach for combining program
slicing and algorithmic debugging for debugging of procedural
programs was introduced. The ideas of using test specifications and
test results can be easily incorporated into our approach. In
contrast to~\cite{fritz91} we can change the debugging performance by
changing the underlying model without changing the underlying
algorithms. 

\begin{figure}[t]
\begin{center}
\epsfig{file=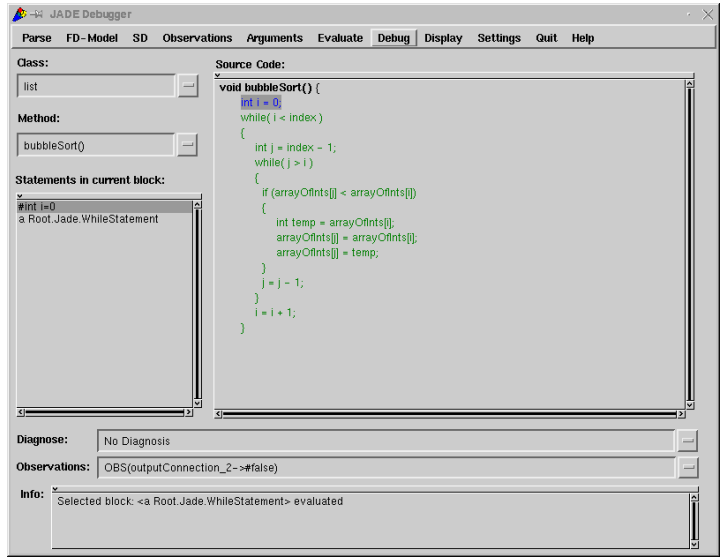,
width=9.5cm}
\end{center}
\caption{The {\sf JADE} debugger main window}
\label{fig:jadeMainWindow}
\end{figure}

The {\sf JADE} debugger is a prototype system for research purposes
and for demonstrating the underlying model-based techniques. Its main
application interface can be seen in Figure~\ref{fig:jadeMainWindow}.
The system is not yet applicable in a real production environment.
The incorporated Java evaluator that is used for computing variable
values to be presented to the user during debugging implements only
the basic Java functionality and some important classes and is far
away from being JDK compliant. In addition, {\sf JADE} assumes that
the Java source code for all involved classes is available.  Both
problems have to be tackled for a production version of {\sf JADE}.

We have described the application of model-based technology in the
building of a prototype intelligent debugger for Java programs.  The
approach is based on automatically building a formal internal model of
the executed program, which a generic diagnosis engine then uses,
together with observations of incorrect program output, for
identifying possible sources of the error in the program. This
combines modeling flexibility with the ability to reuse standard
algorithms like measurement selection.  Following an approach that we
have followed in the domain of hardware design languages, the
currently used model is purely dependency based for simplicity and
quick diagnosis, and will be combined with a more detailed value-based
model for improved discrimination capabilities.

\end{document}